\documentclass[twocolumn,aps,prl]{revtex4}
\usepackage[T1]{fontenc}
\usepackage[latin1]{inputenc}
\usepackage{graphics}

\makeatletter

\providecommand{\LyX}{L\kern-.1667em\lower.25em\hbox{Y}\kern-.125emX\@}

\makeatother
\begin{document}

\title{ORIGIN OF THE VIOLATION OF THE FLUCTUATION-DISSIPATION THEOREM IN
SYSTEMS WITH ACTIVATED DYNAMICS }

\author{A. P\'{e}rez-Madrid\( ^{1} \), D. Reguera\( ^{2} \), and J.M.
Rub\'{\i}\( ^{1} \)}

\address{\( ^{1} \)Departament de F\'{\i}sica Fonamental-CER Física de Sistemes
Complexos, Universitat de Barcelona, Diagonal 647, 08028 Barcelona,
Spain}

\address{\( ^{2} \)Department of Chemistry and Biochemistry, UCLA, 607 Charles
E. Young East Drive, 90095-1569, Los Angeles, USA}

\begin{abstract}
We analyze the validity of the fluctuation-dissipation theorem for
slow relaxation systems \textbf{}in the context of mesoscopic nonequilibrium
thermodynamics. We demonstrate that the violation arises as a natural
consequence of the elimination of fast variables in the description
of a glassy system, and it is intrinsically related to the underlying
activated nature of slow relaxation. In addition, we show that the
concept of effective temperature, introduced to characterize the magnitude
of the violation, is not robust since it is observable-dependent,
can diverge, or even be negative.
\end{abstract}


\maketitle
Many nonequilibrium systems in nature evolve in time following slow
relaxation processes. Examples of this behavior are usually encountered
in glassy systems \cite{sitges}, polymers \cite{angell2}, granular
flows \cite{nagel}, foams \cite{cates}, and crumpled materials \cite{nagel2}
to mention just a few. A complete and satisfactory characterization
of these systems constitutes nowadays one of the most challenging
issues of nonequilibrium statistical physics. The main feature of
slow processes is that the relaxation time may exceed significantly
the observation time scale in such a way that the system can be considered
as being permanently \emph{}out of equilibrium. This peculiarity is
the origin of a distinctive behavior which differs markedly from the
case in which relaxation occurs in shorter time scales. The existence
of aging regimes \cite{kurchan} and the violation of the fluctuation-dissipation
theorem (FDT) \cite{cugliand-pre} constitute examples of this behavior.

For all these reasons, the straightforward application of equilibrium
\textbf{}concepts, appropriate to describe fast relaxation processes,
to out of equilibrium situations, inherent to slow relaxation dynamics,
becomes in principle doubtful. However, our purpose in this Letter
is to show that, when nonequilibrium thermodynamic concepts are applied
at the mesoscopic level \cite{jm}, one may justify many of the peculiarities
of the behavior observed in glassy systems. In particular, we will
show that the violation of FDT \emph{}is a natural consequence of
the activated nature of the dynamics of a slow \textbf{}relaxing \textbf{}system.
Starting from a more detailed description in which the system can
be safely considered as near equilibrium and evolves via a diffusion
process, we will show that the implicit elimination of the fast variables,
leads to an activated regime where the system becomes far from equilibrium
and consequently the FDT is not fulfilled. Coarsening the level of
description is then the origin of the violation of the FDT in strong
glasses. Precisely, one way to characterize this violation is through
the concept of effective temperature. We will also discuss the validity
and robustness of this concept. 

It is well established that the evolution of many systems can be described
in terms of its energy landscape \cite{angell,debene-stilinger},
representing the (free) \textbf{}energy as a function of an order
parameter or reaction coordinate \( \gamma  \) \cite{oppenheim}.
Complex systems exhibit a very intricate landscape with a great multiplicity
of wells separated by barriers. Whereas at high temperatures the system
may explore the whole landscape at low enough temperatures the dynamics
reduces basically to two elementary processes: a fast relaxation toward
the local minima via a diffusion process, and a slow activated process
in which the system overcomes the barrier toward the next minimum.
The presence of the barriers is thus the cause for the slow evolution
of the system. Hence, the case of a single barrier captures the essential
mechanism of the slow dynamics. To show how the activated nature of
the slow evolution of the system can be responsible of some of the
peculiarities of the response of glasses we will then focus on the
simplified model of a bistable potential.

It is then plausible to assume that the evolution of the system occurs
via a diffusion process through its potential landscape parameterized
by the \( \gamma  \)-coordinate, which will be characterized by the
diffusion current \( J(\gamma ,t) \) and the corresponding chemical
potential \( \mu (\gamma ,t) \). As any diffusion process, it can
be treated in the framework of nonequilibrium thermodynamics \cite{degroot}.
Assuming local equilibrium in \( \gamma  \)-space, variations of
the entropy \( \delta s \) related to changes in the probability
density \( \rho (\gamma ,t) \) are given through the Gibbs equation 

\begin{equation}
\label{gibbs}
\delta s=-\frac{1}{T}\int \mu \left( \gamma ,t\right) \delta \rho \left( \gamma ,t\right) d\gamma ,
\end{equation}
 where \( T \) is the temperature. 

The entropy production inherent to the diffusion process, \( \sigma \equiv \partial s/\partial t \),\begin{equation}
\label{entropyproduction}
\sigma =-\frac{1}{T}\int J(\gamma ,t)\frac{\partial }{\partial \gamma }\mu (\gamma ,t)d\gamma ,
\end{equation}
 follows from Eq. (\ref{gibbs}), after using the continuity equation
in \( \gamma  \)-space,  

\begin{equation}
\label{continuity}
\frac{\partial }{\partial t}\rho \left( \gamma ,t\right) =-\frac{\partial }{\partial \gamma }J\left( \gamma ,t\right) .
\end{equation}
 From that expression one then infers the relation between current
and thermodynamic {}``force'' \( J\left( \gamma ,t\right) =-\frac{1}{T}\int L(\gamma ,\gamma ')\frac{\partial }{\partial \gamma '}\mu (\gamma ',t)d\gamma '. \)
The assumptions of locality in \( \gamma  \)-space, for which \( L(\gamma ,\gamma ')=L(\gamma )\delta (\gamma -\gamma ') \),
and ideality, imposing the form of the chemical potential \( \mu \left( \gamma ,t\right) =k_{B}T\ln \rho \left( \gamma ,t\right) +\Phi \left( \gamma \right)  \),
with \( k_{B} \) being the Boltzmann constant, and \( \Phi \left( \gamma \right)  \)
the bistable potential, provide the diffusion current in \( \gamma  \)-space\begin{equation}
\label{current}
J\left( \gamma ,t\right) =-De^{-\Phi /k_{B}T}\frac{\partial }{\partial \gamma }e^{\mu /k_{B}T},
\end{equation}
 where \( D=k_{B}L/\rho  \) is the diffusion coefficient, taken to
be constant as a first approximation. When this phenomenological relation
is substituted into the continuity equation (\ref{continuity}) one
obtains the diffusion equation

\begin{equation}
\label{dif eq}
\frac{\partial }{\partial t}\rho \left( \gamma ,t\right) =\frac{\partial }{\partial \gamma }D\left( \frac{\partial }{\partial \gamma }\rho \left( \gamma ,t\right) +\frac{\rho (\gamma ,t)}{k_{B}T}\frac{\partial }{\partial \gamma }\Phi (\gamma )\right) ,
\end{equation}
 which governs the evolution of the average \emph{}probability density.
This result agrees with the one derived from a master equation \cite{oppenheim},
and indicates that nonequilibrium thermodynamics can be used at mesoscopic
level to provide the fundamental kinetic laws of the Fokker-Planck
type governing the dynamics. 

The probability density is subjected to fluctuations, which may be
introduced through a random contribution to the current, \( J^{r}\left( \gamma ,t\right)  \),
in Eq. (\ref{continuity}) \cite{ignacio}, satisfying the fluctuation-dissipation
theorem in \( \gamma  \)-space\begin{equation}
\label{fdt}
\left\langle J^{r}\left( \gamma ,t\right) J^{r}\left( \gamma ^{\prime },t^{\prime }\right) \right\rangle =2D\left\langle \rho \left( \gamma ,t\right) \right\rangle \delta (\gamma -\gamma ^{\prime })\delta (t-t^{\prime }),
\end{equation}
 where \( \left\langle \rho \left( \gamma ,t\right) \right\rangle  \)
is the solution of Eq. (\ref{dif eq}). The formulation of a \emph{}FDT
is intimately related to the fact that the system is in local equilibrium
in \( \gamma  \)-space. 

When the height of the energy barrier separating the two minima of
the potential is large compared to thermal energy, which happens at
low enough temperatures, a fast relaxation toward the local minima
occurs, and the system achieves a state of quasi-equilibrium characterized
by equilibration in each well\emph{.} The chemical potential then
becomes a piece-wise continuous function of the coordinates, \emph{\( \mu \left( \gamma ,t\right) =\mu \left( \gamma _{1},t\right) \Theta \left( \gamma _{o}-\gamma \right) +\mu \left( \gamma _{2},t\right) \Theta \left( \gamma -\gamma _{o}\right) , \)}
and consequently the probability density achieves the form

\begin{eqnarray}
\rho \left( \gamma ,t\right)  & = & \rho _{1}\left( t\right) e^{-\left\{ \Phi \left( \gamma \right) -\Phi \left( \gamma _{1}\right) \right\} /k_{B}T}\Theta \left( \gamma _{o}-\gamma \right) \label{stationary density} \\
\,  & \: + & \rho _{2}\left( t\right) e^{-\left\{ \Phi \left( \gamma \right) -\Phi \left( \gamma _{2}\right) \right\} /k_{B}T}\Theta \left( \gamma -\gamma _{o}\right) .\nonumber 
\end{eqnarray}
 Here \emph{\( \rho _{1}\left( t\right) \equiv \rho (\gamma _{1},t) \)}
and \emph{\( \rho _{2}\left( t\right) \equiv \rho (\gamma _{2},t) \)}
are the values of the probability density at the minima, \emph{\( \Theta \left( x\right)  \)}
is the unit step function, and \( \gamma _{o} \), \( \gamma _{1} \),
and \( \gamma _{2} \) are the coordinates of the maximum, and the
minima of the potential, respectively. 

Hence, once the fast relaxation toward the local minima has occurred,
the evolution of the system proceeds by jumps from one well to the
other undergoing an activated process. In this situation, a contracted
description performed in terms of the populations at the wells can
be adopted. This description corresponds to \emph{}that of \emph{}the
two level model for a glass \cite{fisher,langer}, a minimal model
which evolves according to an activated dynamics \cite{hanggi} conferring
him the characteristic aging properties of glasses, closely related
to hysteresis \cite{langer2}. Defining the integrated probability
\( N(\gamma ,t)=\int ^{\gamma }_{-\infty }d\gamma '\rho (\gamma ',t), \)
and by integration of the continuity equation (\ref{continuity})
we obtain 

\begin{equation}
\label{integ_cont_eq}
\frac{\partial }{\partial t}N\left( \gamma ,t\right) =-J^{s}\left( \gamma ,t\right) -J^{r}\left( \gamma ,t\right) .
\end{equation}
To proceed with the contraction of the dynamics from the diffusion
regime to the two level regime we will introduce the integral operator
\( P \) acting on a function in \( \gamma  \)-space as 

\begin{equation}
\label{operador_agus}
P\psi (\gamma )=\frac{1}{\int ^{\gamma _{2}}_{\gamma _{1}}\, d\gamma e^{\Phi /k_{B}T}}\int ^{\gamma _{2}}_{\gamma _{1}}\, d\gamma e^{\Phi /k_{B}T}\psi (\gamma ).
\end{equation}
Projecting both sides of Eq. (\ref{integ_cont_eq}) with \( P \),
using Eqs. (\ref{current}) and (\ref{stationary density}), and evaluating
the integrals using the steepest descent approximation, we obtain
the equation governing the dynamics of the two state system\cite{ignacio}\begin{equation}
\label{kinetic equations}
\frac{dn_{1}}{dt}=-\frac{dn_{2}}{dt}=-J(t)-J^{r}(t),
\end{equation}
where \( n_{1}(t)\equiv N(\gamma _{0},t) \) and \( n_{2}(t)=1-n_{1}(t) \)
are the {}``populations'' at each side of the barrier. The value
of the systematic current \( J(t)\equiv PJ\left( \gamma ,t\right)  \),
which is the net current on top of the barrier, \emph{}is given by 

\begin{equation}
\label{j determinista}
J(t)=k_{\rightarrow }n_{1}-k_{\leftarrow }n_{2}\equiv J_{\rightarrow }-J_{\leftarrow },
\end{equation}
 whereas \( J^{r}(t)=PJ^{r}\left( \gamma ,t\right)  \), is the random
current, whose correlation follows from Eq. (\ref{fdt})\begin{equation}
\label{correlaciones j random}
\left\langle J^{r}(t)J^{r}(t^{\prime })\right\rangle =\left( k_{\rightarrow }\left\langle n_{1}\right\rangle +k_{\leftarrow }\left\langle n_{2}\right\rangle \right) \delta \left( t-t'\right) .
\end{equation}
 In the previous expressions, \( k_{\rightarrow } \) and \( k_{\leftarrow } \)
are the forward and backward rate constants \begin{equation}
\label{rate constants}
k_{\rightarrow ,\leftarrow }=\frac{D\sqrt{\Phi'' (\gamma _{1,2})\mid \Phi'' (\gamma _{0})\mid }}{2\pi k_{B}T}\exp \bigg \{\frac{\Phi (\gamma _{1,2})-\Phi (\gamma _{0})}{k_{B}T}\Bigg \}.
\end{equation}
 It is important to highlight that \textbf{}Eq. (\ref{correlaciones j random})
evidences that the fluctuation-dissipation theorem is violated in
the activated process. Only for fluctuations around equilibrium this
equation becomes \( \langle J^{r}(t)J^{r}(t)\rangle =2k_{\rightarrow }n_{1}^{eq}\delta (t-t'), \)

\noindent which is the formulation of the fluctuation-dissipation
theorem\cite{zwanzig libro}. In fact, Eq. (10) with this prescription
constitutes a Orstein-Uhlenbeck process\textbf{.} The failure of the
theorem, which was initially \textbf{}valid in \( \gamma  \)-space,
results from the coarsening of the description. When the dynamical
description is carried out in terms of the reaction coordinate, the
system progressively passes from one state to the other, which makes
it possible the assumption of local equilibrium and the formulation
of a mesoscopic nonequilibrium thermodynamics. \textbf{}However, when
we describe the system in a characteristic time scale similar to the
observation time, we are only capturing the activated process, which
is not near equilibrium and accordingly the FDT does not hold.

The model we have introduced facilitates the analysis of the nonequilibrium
response of the system. Let us consider, for example, the case of
a dynamical observable \( O(\gamma ) \) (energy, density, magnetization,
etc.). Its mean value, in the quasi-stationary regime, is \( \left\langle \delta O(t)\right\rangle =\int O(\gamma )\delta \rho (\gamma ,t)d\gamma =(O_{1}-O_{2})\delta n_{1} \),
where \( \left\langle \delta O(t)\right\rangle =\left\langle O(t)\right\rangle -\left\langle O\right\rangle _{eq}, \)
and \( O_{1} \) and \( O_{2} \) constitute the values of \( O(\gamma ) \)
in the states \( \gamma _{1} \) and \( \gamma _{2} \), respectively.
The response to an external perturbation \( -\epsilon O(\gamma ) \),
plugged in at instant \( t_{w} \), will be characterized by the response
function \( R(t,t_{w})=\left. \partial \left\langle \delta O(t)\right\rangle /\partial \epsilon (t_{w})\right| _{\epsilon \rightarrow 0} \).
This quantity can be calculated from Eq. (\ref{kinetic equations}),
yielding\begin{eqnarray}
R_{O}(t,t_{w})=e^{-(t-t_{w})/\tau }(O_{1}-O_{2}) &  & \nonumber \\
\left[ J_{\rightarrow }(O_{1}-O_{0})-J_{\leftarrow }(O_{2}-O_{0})\right]  & ,\label{respuesta 2} 
\end{eqnarray}
 where \( \tau =\left( k_{\rightarrow }+k_{\leftarrow }\right) ^{-1} \)
is the relaxation time for the activated process, which in view of
Eqs. (\ref{rate constants}) is of the Arrhenius type. From Eqs. (\ref{kinetic equations})
and (\ref{correlaciones j random}) one can also calculate the correlation
function \cite{risken}, \( C_{O}(t,t_{w})=\left\langle \left( O(t)-\left\langle O(t)\right\rangle \right) \left( O(t_{w})-\left\langle O(t_{w})\right\rangle \right) \right\rangle  \),
which for \( t>t_{w} \) and in the limit of large \( t,\; t_{w} \),
is given by\begin{eqnarray}
C_{O}(t,t_{w})=(O_{1}-O_{2})^{2}\tau e^{-(t-t_{w})/\tau } &  & \nonumber \\
\left\{ \left( k_{\rightarrow }-k_{\leftarrow }\right) \left\langle \delta n_{1}(t_{w})\right\rangle +k_{\rightarrow }n_{1}^{eq}\right\} . &  & \label{correlacion} 
\end{eqnarray}

At equilibrium, \emph{i.e. \( J_{\rightarrow }=J_{\leftarrow } \)},
the response reduces to \begin{equation}
\label{respuesta eq}
R^{eq}_{O}(t,t_{w})=(O_{1}-O_{2})^{2}\frac{k_{\rightarrow }n^{eq}_{1}}{k_{B}T}e^{-(t-t_{w})/\tau },
\end{equation}
and is proportional to the time derivative of the equilibrium correlation
obtained from Eq. (\ref{correlacion}), 

\begin{equation}
\label{fdtv}
\frac{\partial C^{eq}_{O}(t,t_{w})}{\partial t_{w}}=(O_{1}-O_{2})^{2}k_{\rightarrow }n^{eq}_{1}e^{-(t-t_{w})/\tau }.
\end{equation}
We then recover the FDT relation \( R^{eq}_{O}=1/k_{B}T\, \partial _{t_{w}}C^{eq}_{O} \),
which holds irrespective the observable we are considering. Out of
equilibrium the FDT is not fulfilled, and its violation is usually
quantified in terms of an effective temperature \cite{cugliand-pre},
\( T_{eff}^{O} \), defined as \textbf{}

\begin{equation}
\label{second correlation}
R_{O}(t,t_{w})\equiv \frac{1}{k_{B}T_{eff}^{O}}\frac{\partial }{\partial t_{w}}C_{O}(t,t_{w}).
\end{equation}
 For the model we are considering, the effective temperature, obtained
from Eqs. (\ref{respuesta 2}) and (\ref{correlacion}), becomes

\begin{equation}
\label{Teff 2}
T_{eff}^{O}=T\frac{1}{Ae^{-t_{w}/\tau }+\left( 1-e^{-t_{w}/\tau }\right) },
\end{equation}
being \( A=\frac{k_{\rightarrow }\left\langle n_{1}(0)\right\rangle \left( O_{1}-O_{0}\right) -k_{\leftarrow }\left\langle n_{2}(0)\right\rangle \left( O_{2}-O_{0}\right) }{k_{\rightarrow }n_{1}^{eq}\left( O_{1}-O_{2}\right) }. \) 

This expression reveals important conclusions. The effective temperature
\( T^{O}_{eff} \) \emph{does} depend on the observable \( O(\gamma ) \)
and explicitly on the waiting time \( t_{w} \). The dependence on
the observable, which has also been found in a trap model for a glass
\cite{sollich} and in experiments \cite{ciliberto-physica D}, evidences
that the effective temperature is not a robust quantity. Only for
small deviation from equilibrium or when \( t_{w}\gg \tau  \), one
recovers the familiar result \( T^{O}_{eff}=T \) for all observables.
It should be noted that our results, obtained by means of a non-mean
field approach, differ from the ones following from \emph{}mean field
models \cite{kurchan, kurchan3} \textbf{}(which yield an effective
temperature independent of the observable) because the latter do not
take into account the activated nature of the dynamics \textbf{}\cite{sollich-jpcm,kawasaki2}.
\textbf{}It also is worth to mention that, since \textbf{\( T^{O}_{eff} \)}
depends on \( t_{w} \), the value of the effective temperature inferred
from the slope of the FD plots, which represent the integrated response
of the system \( \chi (t,t_{w})=\int ^{t}_{tw}dt'R(t,t') \) against
the correlation function, \( C(t,t_{w}) \), it is \emph{not} the
same as the one defined through Eq. (\ref{second correlation}).

\begin{figure}
\resizebox*{0.45\columnwidth}{!}{\includegraphics{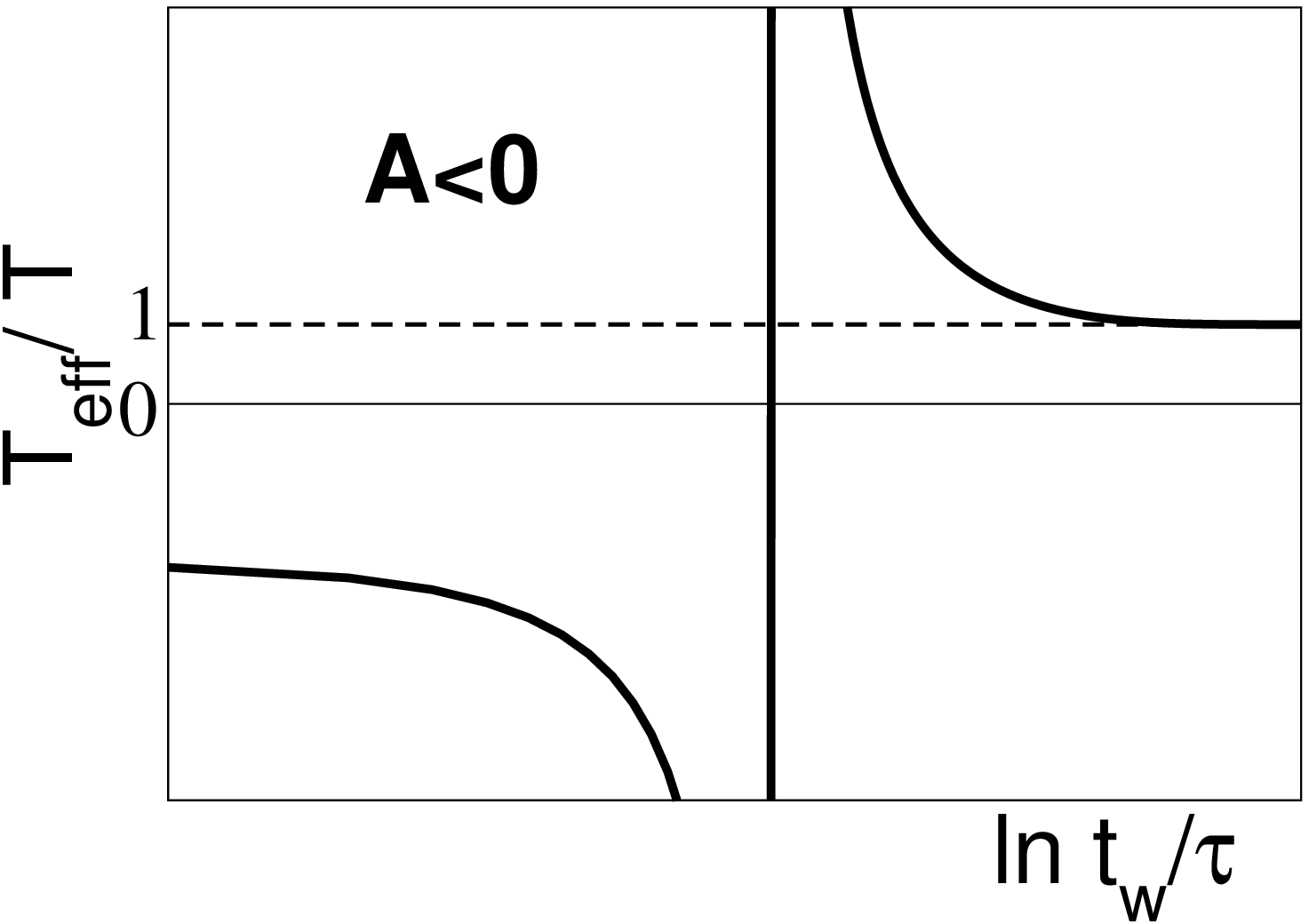}} \hspace{.5cm}\resizebox*{0.45\columnwidth}{!}{\includegraphics{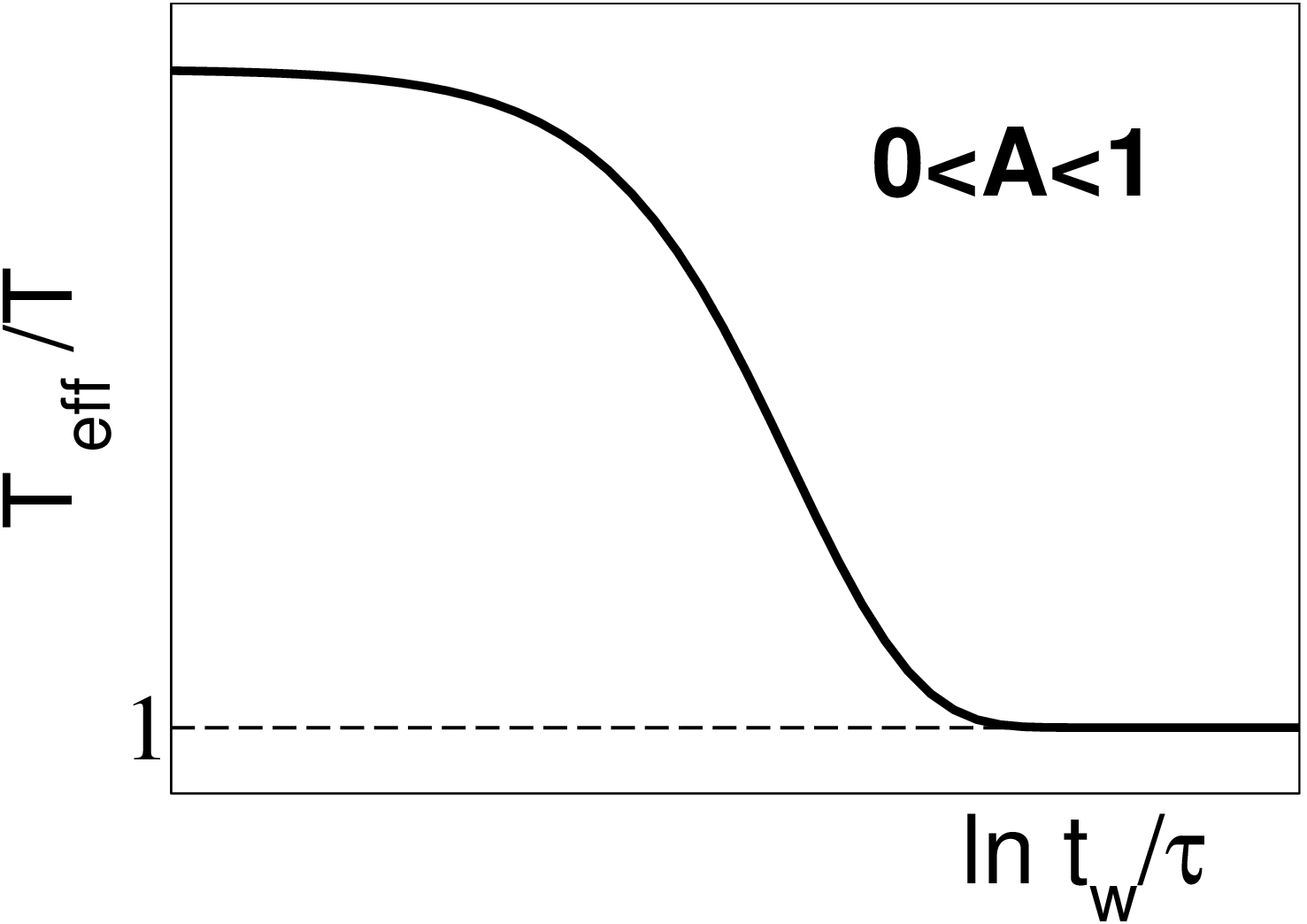}} \vspace{.45cm}

{\centering \resizebox*{0.45\columnwidth}{!}{\includegraphics{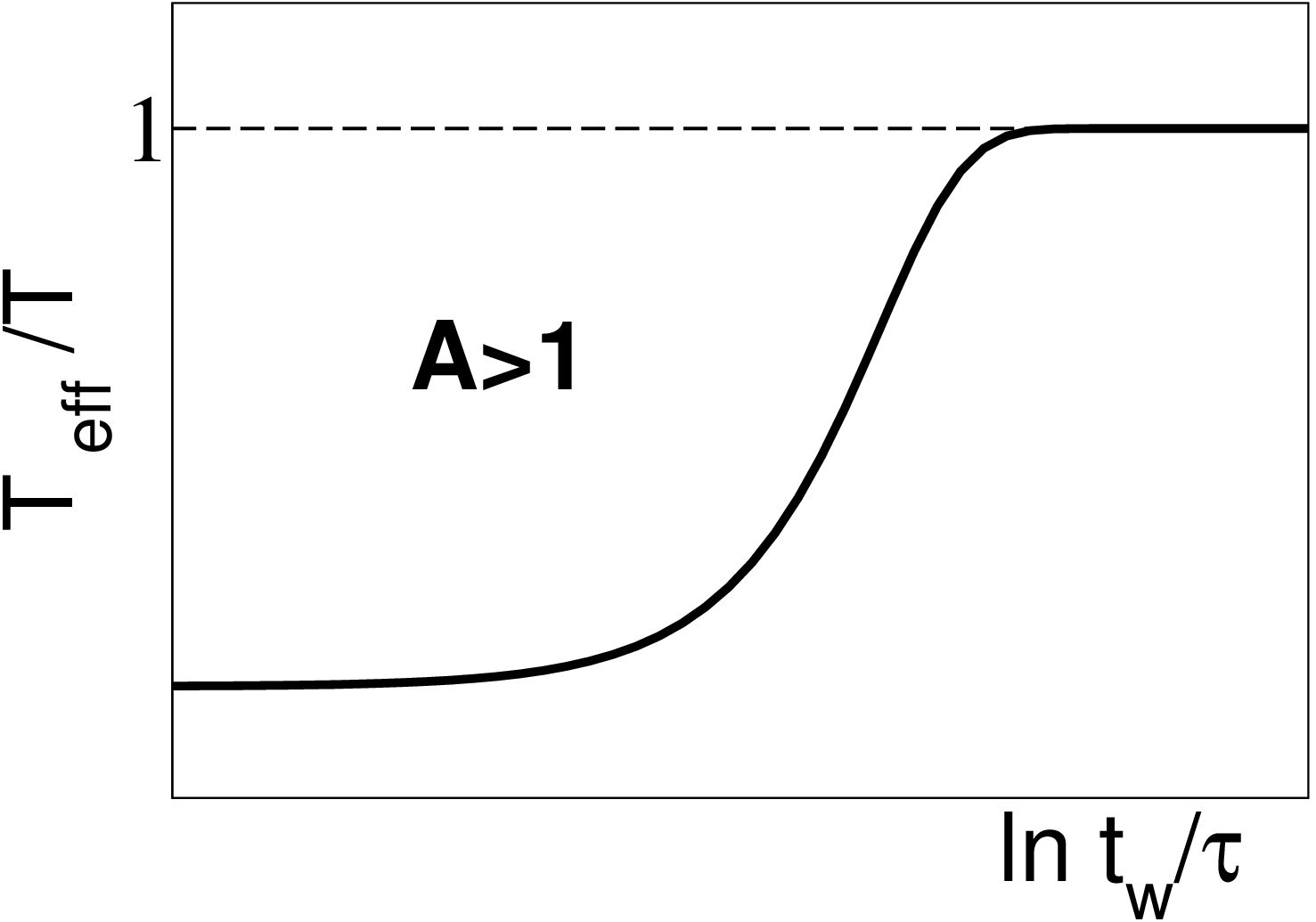}} \par}

\caption{Qualitative behavior of \protect\( T_{eff}/T\protect \) as a function
of \protect\( t_{w}/\tau \protect \) for \protect\( A<0\protect \),
\protect\( 0<A<1\protect \), and \protect\( A>1\protect \). \label{teff de tw}}
\end{figure}

Several interesting behaviors can be identified upon variation of
the parameter \( A \) in Eq. (\ref{Teff 2}). \textbf{}For \textbf{\( 0<A<1 \)},
the effective temperature is higher than the temperature of the bath
\( T, \) in agreement with the experimental measurements reported
in \cite{israeloff}. Contrarily, if \( A>1 \) the effective temperature
is lower than the bath temperature \( T \) ; whereas, if \( A<0 \),
\( T^{O}_{eff} \) may diverge as predicted in \cite{cugliand-pre},
numerically verified in \cite{barrat}, and experimentally suggested
in \cite{ciliberto}, or even become negative \cite{sollich-jpcm}.
All these cases are illustrated in Fig. \ref{teff de tw}, \textbf{}and
arise from the peculiar behavior of the nonequilibrium response of
an activated process. The effective temperature is essentially a measure
of the ratio between the equilibrium and the nonequilibrium responses
of the system. When this ratio is smaller (larger) than one, then
\( T^{O}_{eff}<(>)T \). A divergence in \( T^{O}_{eff} \) occurs
when the nonequilibrium response vanishes, and {}``negative'' effective
temperatures would be caused by nonequilibrium responses having a
different sign than its equilibrium counterpart. These anomalous behaviors
can be tuned by a proper choice of initial conditions and observables.
\textbf{}To illustrate that fact, \textbf{}we have implemented our
theory for two \textbf{}examples of bistable potentials\textbf{:}
a \textbf{}quartic potential \( V(\gamma )=\gamma ^{4}/4+a\gamma ^{3}/3-\gamma ^{2}/2-a\gamma  \)
, being \( a \) an adjustable parameter responsible for its asymmetry\textbf{,}
and \textbf{}the potential \textbf{\( V(\gamma )=-\varepsilon \cos (\gamma )+1/2\sin ^{2}(\gamma ) \),}
describing a monodomain magnetic particle \cite{agusti2}. Selecting
different observables and initial conditions, we obtain in both cases
the behaviors of the effective temperature shown in Fig. \ref{teff de tw}. 
In summary, we have shown that the origin of the violation of the
FDT is the drastic elimination of variables one tacitly performs to
model the system in the experimental time scale. \textbf{}At this
level, the system evolves undergoing an activated dynamics, requiring
big amounts of energy to surmount the barriers. Consequently, it is
always far from equilibrium and the FDT, a result strictly valid at
or near equilibrium, is not \textbf{}fulfilled\textbf{.} In the more
complete scenario, when instead of jumping between two states the
system reaches a different state passing progressively from intermediate
configurations, \emph{i.e.} diffusing in a configuration space, local
equilibrium can be established\textbf{.} One can then proceed with
the formulation of a mesoscopic nonequilibrium thermodynamics \cite{david},
perfectly compatible with the Fokker-Planck level of description\textbf{,}
whose underlying stochastic kinetics satisfies FDT. A widely-used
way of quantifying the FDT violation is through the definition of
an effective temperature. Our analysis shows that this concept suffers
from a lack of robustness, since its value depends on the dynamical
variable we measure, and can diverge or even become negative. All
these problems limit the scope and question the usefulness of this
quantity in the description of glassy systems where the activated
dynamics is an unavoidable ingredient. \textbf{}

The theory we have developed provides a useful framework to describe
the behavior of systems with slow dynamic bridging the macroscopic
and the mesoscopic descriptions, by indicating the way to generalize
local equilibrium concepts.

\begin{acknowledgments}
This work has been partially supported by DGICYT of the Spanish Government
under grant PB98-1258, and by the NSF \textbf{}grant No. CHE-0076384\textbf{.}
We gratefully acknowledge interesting discussions with D. Bedeaux,
S. Kjelstrup, H. Reiss, and J.M. Vilar.
\end{acknowledgments}

\end{document}